\documentclass[conference]{IEEEtran}
\usepackage{cite}
\usepackage{amsmath,amssymb,amsfonts}
\usepackage{algorithmic}
\usepackage{graphicx}
\usepackage{textcomp}
\usepackage{xcolor}
\usepackage{float}
\usepackage{url}
\usepackage{caption}
\def\BibTeX{{\rm B\kern-.05em{\sc i\kern-.025em b}\kern-.08em
    T\kern-.1667em\lower.7ex\hbox{E}\kern-.125emX}}

\begin{document}

\title{How Gamification Affects Software Developers:\\ Cautionary Evidence from a Natural Experiment on GitHub}

\author{\IEEEauthorblockN{Lukas Moldon}
\IEEEauthorblockA{RWTH Aachen University\\
Aachen, Germany}
\and
\IEEEauthorblockN{Markus Strohmaier}
\IEEEauthorblockA{RWTH Aachen University \&\\
GESIS-Leibniz Institute for the Social Sciences\\
Cologne, Germany\\
0000-0002-5485-5720}
\and
\IEEEauthorblockN{Johannes Wachs}
\IEEEauthorblockA{Vienna Uni. of Econ. and Business \&\\
Complexity Science Hub Vienna\\
Vienna, Austria\\
0000-0002-9044-2018\\
johannes.wachs@wu.ac.at}
}

\maketitle

\thispagestyle{plain}
\pagestyle{plain}

\begin{abstract}
We examine how the behavior of software developers changes in response to removing gamification elements from GitHub, an online platform for collaborative programming and software development. We find that the unannounced removal of daily activity streak counters from the user interface (from user profile pages) was followed by significant changes in behavior. Long-running streaks of activity were abandoned and became less common. Weekend activity decreased and days in which developers made a single contribution became less common. Synchronization of streaking behavior in the platform's social network also decreased, suggesting that gamification is a powerful channel for social influence. Focusing on a set of software developers that were publicly pursuing a goal to make contributions for 100 days in a row, we find that some of these developers abandon this quest following the removal of the public streak counter. Our findings provide evidence for the significant impact of gamification on the behavior of developers on large collaborative programming and software development platforms. They urge caution: gamification can steer the behavior of software developers in unexpected and unwanted directions.
\end{abstract}


\section{Introduction}

Online platforms often employ gamification elements to increase user participation and to steer user behavior in desired directions. Points, badges, and leaderboards are known to encourage people to spend more time interacting with a system~\cite{hamari2014does}. These elements also play an important role in building user reputation and trust in a community. Points and tokens users earn can have value beyond the platform in question, for instance as credentials in the labor market. A good gamification system can grow user engagement and increase social interaction and collaboration. However gamification can also steer users astray by promoting narrowly defined goals and encouraging unreasonable levels of activity. These potential downsides present especially pressing problems when the platform has significant social and economic implications for its users.

It is important to understand the influence of gamification on user behavior because it has spread to all corners of the web. Ecommerce sellers on sites like eBay collect references to signal their trustworthiness~\cite{chou2019actionable}. Freelance workers covering a wide range of industries from digital design to food delivery display badges of accomplishments on their personal profiles~\cite{hannak2017bias,schorpf2017triangular}. Gamification is also used by governments to nudge their citizens towards better decisions~\cite{bista2014gamification} and by educators to guide their students to better learning outcomes~\cite{de2016effectiveness}. Yet gamification is no silver bullet: studies have shown that poorly designed games can sap motivation~\cite{yamakami2013gamification,hanus2015assessing} and reorient effort toward chasing metrics rather than substantive outcomes~\cite{chou2019actionable}. When games are used to rank people at work, the high stakes can lead to overwork and interpersonal conflict~\cite{sharone2004engineering}.

Gamification is especially prevalent on platforms used by software engineers for collaborative work~\cite{pedreira2015gamification}. Two distinguished examples are Stack Overflow~\cite{badge}, a large Q\&A community for programming related knowledge, and GitHub~\cite{repobadges}, the largest forum for collaboration in open source software. Gamification plays an important role in open source software because its tradition of decentralized, online collaboration~\cite{crowston2005social} creates a demand for ways to effectively signal commitment, competence, and trustworthiness~\cite{vasilescu2014human}.

The promise of gamification on online platforms in general, and for open source software communities in particular, then, is to increase participation and trust among users. It accomplishes this by rewarding particular kinds of actions and highlighting milestones and successes of a user's career. A vast literature of observational~\cite{nacke2017maturing} and experimental~\cite{hamari2017badges} studies suggests that gamification works in a narrow sense: users respond to these rewards by changing their behavior~\cite{hamari2014does}, and revert to previous patterns when gamification is removed~\cite{thom2012removing}. But as we have noted, just because gamification does steer user behavior, does not mean that the resulting behavior is desirable. Nor does gamification work the same way for everyone: some individuals may genuinely enjoy gamification, but others may ``feel a compulsion [to participate] when the system pulls on psychological levers such as social comparison or rewards''~\cite{kim2016more}. In general the negative effects of gamification elements are understudied in the literature~\cite{hyrynsalmi2017dark}, especially among software developers~\cite{pedreira2015gamification}.

In this paper we demonstrate the behavioral affects of gamification on software developers by studying individuals contributing to GitHub and a natural experiment involving the design of the platform. In May 2016 GitHub removed, without warning or official announcement, two counters from developer profiles that tracked their current and all-time longest streaks of uninterrupted daily contributions. As the change was exogenous, the change in behavioral traces of developers across this change contains more precise information about their relation to gamification than one can typically capture with observational studies~\cite{malik2016identifying,dev2019quantifying}. And because it happened ``live'', on a platform used by hundreds of thousands of people every day, these insights are likely more generalizable than those derived from lab experiments.

We use this setting to test the following research questions relating developer behavior and gamification.

\begin{itemize}
    \item RQ1: Did developer streaking behavior change significantly after the design change?
    \item RQ2: Did the timing and distribution of developer activity change?
    \item RQ3: Did developers use the counters to set and achieve personal goals?
    \item RQ4: Was there a significant change in the correlation of streaking behavior in the social network?
\end{itemize}

These questions serve as a framework to evaluate how gamified streak counters on GitHub affected developer behavior. They also help us diagnose whether or not the counters were effective, both in the sense that they fostered certain kinds of behavior and whether that behavior was, in fact, desirable. To address them, we compiled a database of developers active around the site design change. We observe their activities overtime to record the lengths of uninterrupted streaks of daily activity. We analyze the distribution of these streak lengths and activity in general across the change using a variety of methods. This approach exploits the idea that sudden changes in activity patterns related to streaking in the aftermath of the design change are highly suggestive of gamified behavior.

Our analysis suggests that the removal of the counters was followed by several changes in developer behavior. First we document that many long streaks ongoing at the time of the change are abandoned. In the long term, there are significantly fewer long streaks. This overall change in behavior manifested in particular ways that suggest that the counters were steering behavior in undesirable ways. For instance, developer activity decreased on weekends compared to weekdays, suggesting that the counters were pushing developers to contribute on days they would have otherwise rested. We also find that developers were less likely to make a single contribution in a day after the change, suggesting that developers had previously been consciously maintaining their counters. We speculate that contributions made for the sake of a streak do not represent highly productive work. Finally, we find that the tendency for neighbors in the social network to synchronize in their streaking behavior fell significantly after the change. This suggests that developers were pulled to maintain streaks by peer effects.

These findings provide insight into how gamification changes developer behavior on an important online platform, especially in potentially negative ways. For example, streak-chasing behavior likely had unhealthy externalities on the quality of developer outputs - evidenced by the phenomenon of single contribution days. Though GitHub has removed this particular feature, the lessons we can learn from this particular gamification design can help platform owners design better features in the future.

\section{Background}
In this section we review related work on gamification. We introduce some general findings about the effectiveness of gamification and the different ways it steers human behavior. We then discuss previous work on gamification in the context of computer programming and software development.

\subsection{Gamification and Motivation}
Gamification seems to appear wherever people have meaningful social or economic interactions online. Some kinds of gamification, for example feedback ratings or reputation points, can help grow trust in a community~\cite{basten2017gamification}. Gamification is also used by platforms to increase the frequency, duration, and intensity of user engagement~\cite{rodrigues2016playing,looyestyn2017does}. These goals can be applied to virtuous ends, for example improving educational outcomes among students~\cite{da2016effectiveness}, but can also lead to negative outcomes. It can misdirect effort and incent dishonesty~\cite{welsh2014dark} or lead to overwork or burnout~\cite{andrade2016bright}. Gamification can commodify labor by facilitating the monitoring of workers~\cite{mason2018high}. 

Different implementations of gamification have varying effects on user behavior. Some of this heterogeneity comes from the design of the gamification element in question. For instance \textit{leaderboards}, which publicly rank users over the course of a project or task, seem to drive competitive behavior as relative performance becomes more important~\cite{landers2017gamification}. Users tend to temporarily alter their behavior in order to collect \textit{badges}, tokens which users can display after completing a specified activity~\cite{badge,fanfarelli2015understanding}. On many platforms badges are valuable both for their signal that a user has accomplished a feat, and to grant the user special privileges. 

An important recognition is that not all users are equally interested in engaging with gamified elements. In one study on a platform some users are eager to collect points, others are happy with a more moderate scores, while others are totally uninterested~\cite{eickhoff2012quality}. Several papers have shown that social and cultural effects influence an individual's propensity to respond to gamification~\cite{vasilescu2015gender,almarshedi2017gamification}. 

Previous works explain the adoption of gamified elements by users by probing how they activate or amplify psychological motivations. Some users chase gamified elements for their value as a signal. Others use gamified elements to set goals, either in relation to their own previous outcomes~\cite{anderson2018personal} or in competition with others~\cite{landers2017gamification}. Some people may pursue points to imitate others~\cite{hamari2015working}. Several studies show how gamification can exploit motivations in ways that lead to bad outcomes~\cite{hanus2015assessing,kim2016more}.

\begin{figure*}
\centerline{\includegraphics[width=.9\textwidth]{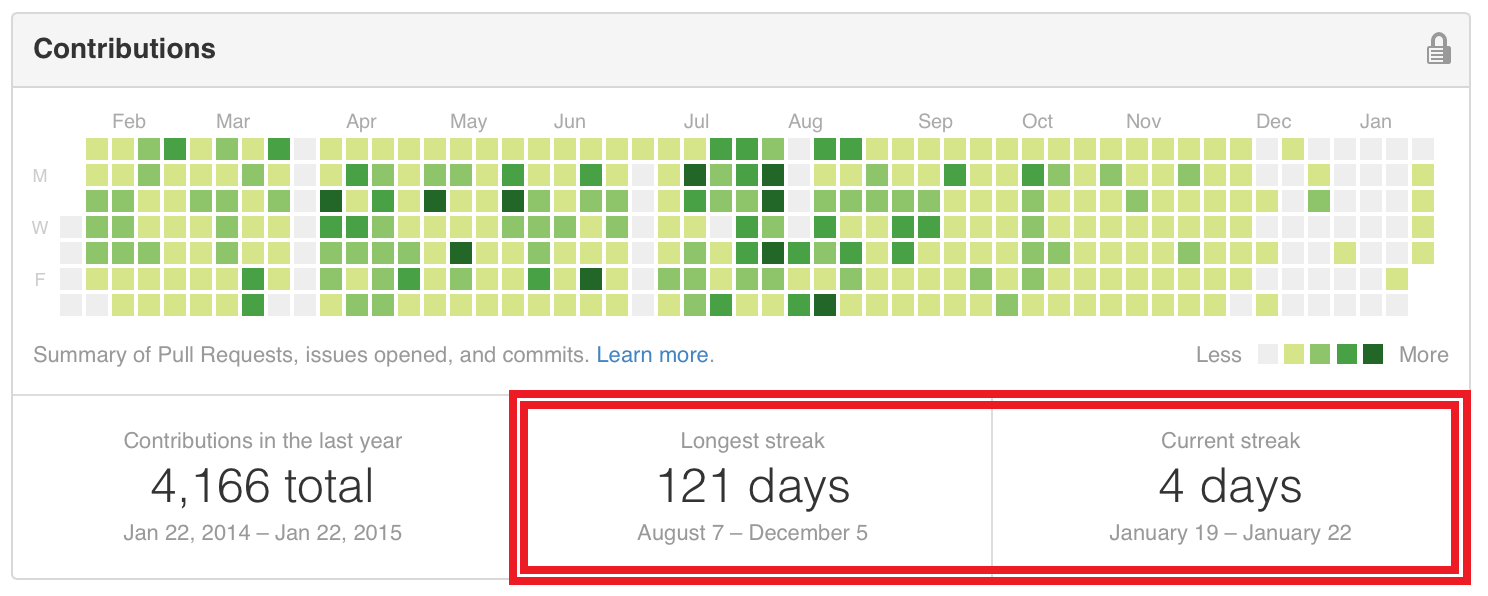}}
\caption{Example of a GitHub user profile's activity data, prior to May 19th, 2016. On that date an unannounced design change removed the two highlighted counters, tracking the lengths of the developer's longest and current streaks of daily activity. Source: \protect\url{https://zachholman.com/posts/streaks/}}
\label{fig:exampleStreakCounter}
\end{figure*}

\subsection{Gamification in Software Development}
As mentioned in the introduction, online communities relating to software development and computer programming tend to have a significant gamification footprint. Collaborative work on software naturally takes place in an online context and the open source software community in particular is highly geographically dispersed~\cite{gonzalez2008geographic,robles2014floss}. As a result, a significant share of interactions takes place on social platforms~\cite{storey2014r}. Trust and reputation are important in these contexts, suggesting that gamified elements have a significant role to play in this community. 

As software development is a quickly changing labor market, non-standard credentials are often used to evaluate job candidates and potential collaborators. For example, instead of first considering an individual's employment history, college degrees, or self-described programming language experience, a hiring manager may prefer to check out an individual's profile on a platform like GitHub~\cite{marlow2015effects}. The signals sent by a few key markers on these profiles can make a big first impression. Indeed, eye-tracking experiments confirm that visitors to a new profile page dwell on counters, badges, and statistics~\cite{ford2019beyond}. In this way, the way in which a developer's history of contributions is represented can have significant impact on how they are evaluated in the future. 

The above-mentioned potential negative side-effects of gamification are especially important in the software development community. On the one hand, open source software communities are widely used throughout the digital economy, often in mission-critical settings~\cite{eghbal2016roads}. On the other, the people in this community are often working long hours, multitasking between many projects~\cite{multitasking}, and are highly stressed~\cite{raman2020stress}, sometimes leading to burnout and project abandonment~\cite{miller2019people}. Because it is difficult to evaluate the quality of contributions to software projects in general, gamification can only set goals that proxy for quality in this context. This presents the risk that gamified software developers chase metrics or optimize behavior in ways that correlate with but do not cause good outcomes. It is imperative that the research community better understands the extent to which gamification can cause harm.

One example of a platform relating to software development with a significant gamification aspect is Stack Overflow, the largest Q\&A platform for questions about computer programming. Several previous studies describe how Stack Overflow users engage with gamification. For instance, users will significantly change their behavior when they are close to obtaining a so-called threshold badge~\cite{badge}, returning to old habits soon after. Over the course of its history, Stack Overflow has introduced several new badges - often leading to sudden changes in behavior observable at the macro scale as users chase these new tokens~\cite{bornfeld2017gamifying}.

\section{Data}
In this section we describe GitHub's gamified elements and the sudden removal of one of these elements in 2016. We then describe how we collected, filtered, and processed the data for the purposes of our analysis. The data and code used to perform this filtering are available at \url{https://github.com/lukasmoldon/GHStreaksThesis}.

\subsection{Gamification on GitHub}
GitHub has had several gamification elements on its site. As of 2020, developer profiles are still adorned with a contribution calendar - a visual representation of the daily intensity of their activity on the site in the last year. Previously GitHub included two counters below each developer's calendar, one reporting the developer's all-time longest streak of consecutive days making a contribution on GitHub, and the other reporting the developer's ongoing streak. We share an example in Figure~\ref{fig:exampleStreakCounter}.

This external shock, which we interpret as a natural-experimental perturbation of gamification on GitHub, serves as the lynch-pin of our analysis. We study changes in behavior relating to streaks around this date assuming that they are at least partially made in response to the removal of the counters. For instance, we will soon observe that there was a significant drop in the number of long streaks that were active shortly following the design change. We interpret this as a response: developers suddenly lost an incentive to maintain their streaks and adjusted their behavior in response.

Besides the counters and the still-present activity calendar, we note that GitHub also has gamification elements in the form of badges for projects. Previous work has demonstrated that project owners seek out these badges, and that projects in turn are evaluated more favorably when they have them~\cite{repobadges}.

\subsection{Data processing}

Our primary data source for GitHub data is GHTorrent, an updating database of information retrieved from the GitHub REST API~\cite{gousios2014lean}. We access data from the June 2019 dump. The data set contains 32.5 million developers, 125 million projects, 100 million opened issues and 1.368 billion unique commits. To  address our research questions we proceeded to filter and modify the data.

\begin{figure*}
\centerline{\includegraphics[width=.9\textwidth]{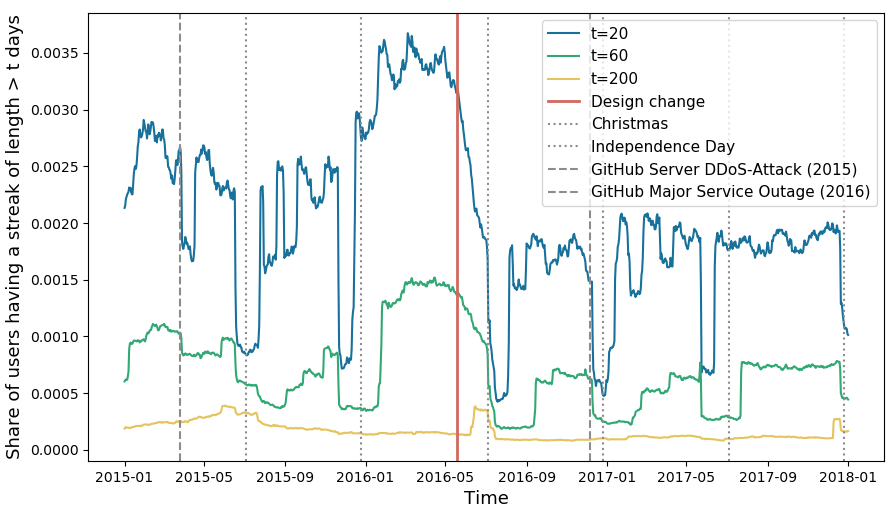}}
\caption{Share of developers having a streak of length $>$ t days for t $\in \{20,60,200\}$: One of the largest drops occurs right after the streak counters were removed from GitHub (red line). Developers tend to abandon their streaks across holidays season (dotted lines). Server outages also influence streaks. (dashed lines).}
\label{fig:shareStreaking}
\end{figure*}

As our primary focus is the platform's design change, we only consider developers who were active on the site around the time of the change. From the original population, we discarded all developers who did not have a commit in a non-forked repository (17.3 million remaining). We also removed developers with more than 100 invalid commit timestamps to filter developers who may have manipulated their activity histories and bot accounts. We consider a timestamp as invalid if it has an illogical format or is unrealistic (i.e. before GitHub was founded or after the data dump was created). Also bots make a significant number of contributions, so it is important to filter them out carefully. Our end sample excludes over 99\% of the bots identified in a recent paper on GitHub bot detection, which we discovered after our analyses were completed~\cite{dey2020detecting}. We kept the remaining developers who had at least 100 commits, were assigned a ``USR'' type (excluding organization accounts), and who had an associated geolocation from GHTorrent, leaving 433,138 developers. We focus on developers which are geolocated for a technical reason. To accurately track daily activity streaks, it is necessary to know a developer's timezone as every timestamp is converted and saved in UTC-0 by GitHub, but streaks are evaluated by local time zones. For example, without knowing that a developer lives in San Francisco, their commit at 8PM local time on a Monday would be incorrectly evaluated as a commit on a Tuesday (3AM in UTC-0). These coordinates are inferred by GHTorrent, using the location free text field on developer profiles and the OpenStreetMap API. More than 85\% of developers in this population joined GitHub before the design change. 

Next we proceeded to tabulate the daily activities of each developer in order to recreate the streak counters they had on each day. Three kinds of contributions counted towards streaks: commits, pull requests and issues. There were some specific rules for these activities to count~\footnote{For a complete list and description see: \url{https://help.GitHub.com/en/articles/why-are-my-contributions-not-showing-up-on-my-profile} (September 28, 2019)}. For instance, contributions had to be associated with a standalone (non-forked) project. For pull requests and issues, we checked if they were made in a forked repository and filter such activity out. However, because 48 million projects represent a forked copy of a corresponding standalone project, commits are assigned to projects 6.252 billion times. Whenever a project is forked, all commits of this origin standalone project are duplicated and assigned to the forked project copy, too. In this case we had to discard commits to forks which were never merged back to the original project. We created filtered databases for each contribution type for all observed developers.

The remaining dataset consists of 433,138 developers with over 290 million valid contributions (including 12.8 million issues and pull requests). In the last step, all contributions were sorted by time and developer. We computed the resulting data set of streaks (start, end) assigned to the corresponding developer ID.

\section{Analysis}

Our overarching empirical strategy is to describe how developer behavior differs across the design change. We view the removal of the counters as a shock: developers did not anticipate this change. As a result, we interpret changes in developer activity relating to streaking across the change as evidence for the effect of gamification on behavior.

\subsection{RQ1: Changes in Developer Behavior}

We first address the question of whether or not we can observe significant changes in overall behavior. We begin by introducing general findings about the share of streaking developers over time. Afterwards we consider the distributions of streaks starting on Mondays and compare the lengths of such streaks before and after the design change.

We calculated the share of all observed developers having a streak with a minimum length of 20, 60 and 200 days for each day. Note that to compute these values we count streaks for each group ($t=$20, 50, 200) from the day they passed the threshold $t$ not from the day they were started. This emphasizes the impact of events such as holidays or service outages. As we are calculating the share of developers with an ongoing streak, we divide the count of such developers by the number of developers in our dataset who had registered at least $t-1$ days before. This adjusts for the growing population of GitHub developers over time. The resulting plot (Figure~\ref{fig:shareStreaking}) shows that the largest drop of streaking developers within 3 years happened immediately after the 19th of March in 2016. Moreover, we observe a long-term decline in the share of developers on long streaks. However, developers having streaks longer than 200 days did not change their behavior directly after the change.

We can make two qualitative interpretations from this analysis. First, we note that many of the sudden drops in streaking behavior around holidays or outages witness a subsequent, quite symmetric recovery roughly $t$ days later. This suggests that there is some natural base rate of streaking. The second is that the decline in streaking in the immediate aftermath of the design change is somewhat more gradual than the other major declines observed around major holidays or platform outages. We interpret this as a population of developers giving up their streaks gradually in response to the design change.

\begin{figure}
\centerline{\includegraphics[width=.5\textwidth]{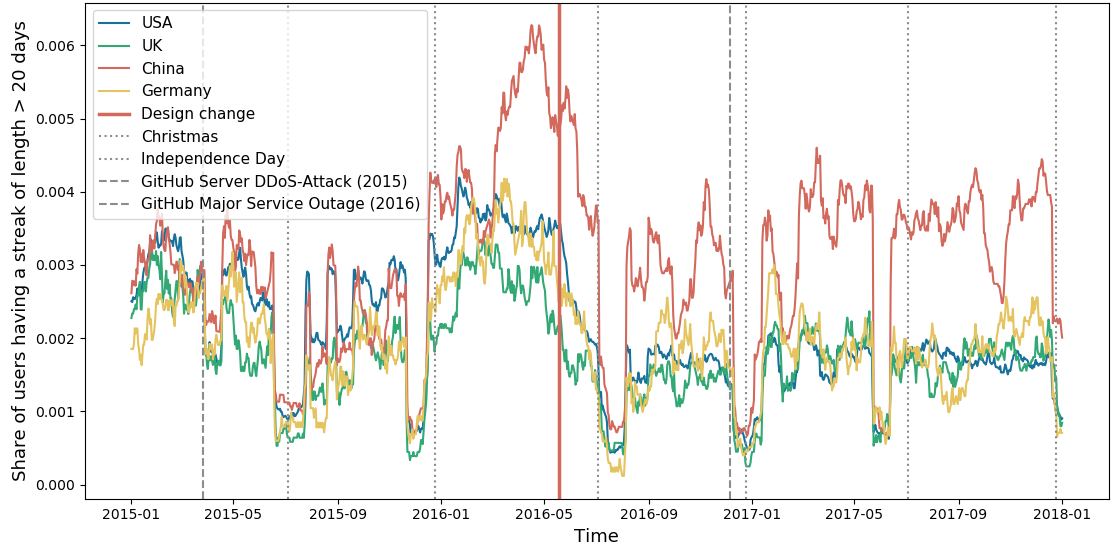}}
\caption{Share of developers from different countries having a streak of length $>$ 20 days: While western countries are affected equally by the design change, developers from China continue streaking afterwards on a similar level.}
\label{fig:shareStreakingCountries}
\end{figure}

Figure~\ref{fig:shareStreakingCountries} focuses on the share of developers with an ongoing streak of length $t=20$, with developers broken up by their country of origin. We use the same geolocation of developers which we use to make time-zone adjustments. We find that while western countries are equally affected by the design change, Chinese developers continue streaking on a similar level (with one temporary drop across the US Independence Day). One explanation could be that Chinese developers often have significantly more demanding working hours than their counterparts in the Western world. This interpretation is supported by recent protests against long working hours on GitHub by Chinese developers and their supporters~\cite{xiaotian996}. We also note that the intensity of the decline in streaks across the design change is similar for all four countries.

To test the statistical significance of the change in share of developers streaking, we zoom in on activity right around the change. We compare the lengths of streaks starting exactly three weeks before the design change with those starting exactly three weeks after the change. A Mann-Whitney-U test indicates that the former collection of streak lengths has a significantly higher average, significant at $p<.01$. We then focus on Mondays to compare the characteristic lengths of streaks around the design change because of the well-documented ``Monday Effect''~\cite{ghtorrent}, which notes that a significant amount of contributions to GitHub take place on Mondays. We report characteristic streak lengths on various Mondays around the removal of the counters in Table~\ref{tab:mondaystreaks}, the likelihood that they last more than a week, as well as odds of a streak lasting more than 14 days conditional on reaching a 7 days. 

\begin{table}[b]
%\hskip-4.3cm
\begin{tabular}{|l|l|l|l|}
\hline
\textbf{Starting date} & \textbf{Avg length} & \textbf{P(len \textgreater 7)} &
\textbf{P(len \textgreater 14 $|$ len \textgreater 7)} \\ \hline
2016/04/18 & 2.38 & 0.52\% & 15\% \\ \hline
2016/04/25 & 2.29 & 0.40\% & 10\% \\ \hline
2016/05/02 & 2.24 & 0.40\% & 11\% \\ \hline
2016/05/09 & 2.36 & 0.43\% & 7\% \\ \hline
2016/05/16 & 2.33 & 0.45\% & 9\% \\ \hline
Change & - & -  & -  \\ \hline
2016/05/23 & 2.30 & 0.39\% & 9\% \\ \hline
2016/05/30 & 2.27 & 0.40\% & 6\% \\ \hline
2016/06/06 & 2.27 & 0.31\% & 5\% \\ \hline
2016/06/13 & 2.24 & 0.27\% & 1\% \\ \hline
2016/06/20 & 2.28 & 0.35\% & 3\% \\ \hline
\end{tabular}
\caption{Comparison of streaks starting on various Mondays around the site design change. The sharper decrease in the probability of long streaks suggests a loss of interest in behavior tracked by the counters removed in the change.}
\label{tab:mondaystreaks}
\end{table}

\begin{figure}[t]
\centerline{\includegraphics[width=0.5\textwidth]{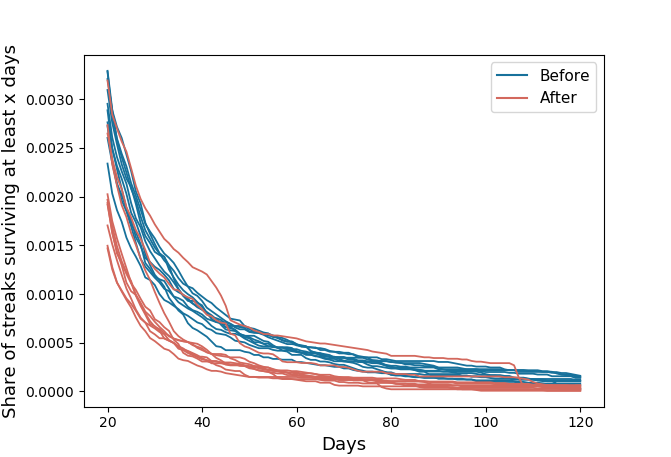}}
\caption{Share of streaks surviving at least x days. Each line represents the survival curve of streaks started in one of the first ten weeks of either 2016 (blue) or 2017 (red). We observe a clear separation - the lower position of the red curves indicates that in 2017, after the site design change, long streaks became less common.}
\label{fig:survival}
\end{figure}

Considering the change in streaking behavior in the long run, we compare all streaks beginning on Mondays of the first ten weeks in 2016, before the design change, with those from 2017. We plot the truncated survival curves of streaks in Figure~\ref{fig:survival}. These curves describe the chance that a streak starting in a given weeks survives $t$ days. The clear separation of most red curves, representing weeks in early 2017, from the blue curves, representing the first weeks from 2016, suggests that a change has taken place. Indeed, we note that the average length of a streak exceeding 14 days declined from nearly 26 days in early 2016 to 22 days in early 2017. At extreme values the change is even more drastic: among streaks of length at least 14, those started in early 2016 were more than twice as likely to exceed 100 days (4.4\%) than those started in early 2017 (2.0\%).

In summary, we found evidence that long streaks were abandoned following the design change, and that new streaks became significantly less common. The difference in frequency of streaks becomes larger as we consider longer streaks. We see evidence for this effect when we zoom in on the weeks around the design change, and when we compare activity across one year.

\subsection{RQ2: Changes in Timing and Distribution of Activity}
Having demonstrated that there is a significant change in streaking behavior following the design change, we now turn to our second question, asking whether the timing and distribution of developer activity changed. We consider two ways in which developer activity may have changed qualitatively. The first is that developers may be more likely to take a break from the platform on weekends. We find evidence of a small but significant drop in the relative share of activity on the weekends. The second is that developers no longer have incentive to make contributions for the sake of extending an ongoing streak. We also find evidence for this phenomenon.

\begin{table*}
\begin{tabular}{l|l|l|l|l|l|l|}
\cline{2-7}
 & \multicolumn{2}{l|}{\textbf{All Developers}} & \multicolumn{2}{l|}{\textbf{With Streak $\ge 20$}} & \multicolumn{2}{l|}{\textbf{With Streak $\ge 30$}} \\ \cline{2-7} 
 & \textbf{Before} & \textbf{After} & \textbf{Before} & \textbf{After} & \textbf{Before} & \textbf{After} \\ \hline
\multicolumn{1}{|l|}{\textbf{Share of contrib. on weekend}} & 0.2188 & 0.2179 & 0.2433 & 0.2399 & 0.2487 & 0.2459 \\ \hline
\multicolumn{1}{|l|}{\textbf{Contrib. on weekends (millions)}} & 45.3 & 48.6 & 13.5 & 10.3 & 9.5 & 7.6 \\ \hline
\end{tabular}
\caption{Share and total amount of weekend activity for all developers and only streaking developers (achieving a streak of length 20 or 30 in the respective time interval) in the year before (B) and after (A) the change. The share of weekend work decreases especially for streaking developers.}
\label{tab:weekendBroader}
\end{table*}

\subsubsection{Weekend Activity}
Even though a significant share of open source development activity occurs on nights and weekends~\cite{claes2018programmers}, weekends are considered a time to rest and spend with friends and family in most cultures around the world. Moreover, sociologists have documented that time is a networked good~\cite{young2014time}, meaning that time for work or recreation is more valuable when it is synchronized with the time of others. Without the incentive to extend a long ongoing streak, we argue that developers will be more likely to take time off on the weekends.

We present descriptive statistics of the relative share of developer contributions before and after the change in Table~\ref{tab:weekendBroader}. We see that the share of activity on the weekend drops among all developers drops (~.09\%), and moreso for those developers who achieve long streaks (.28-.34\%). To test the statistical significance of this change, we build a model.

In the following we focus on active developers with at least 30 contributions in the respective time interval. To test for statistical significance of the design change on weekend work, we apply the regression discontinuity design method~\cite{imbens2008regression}. Our goal is to fit a linear model on the share of weekend activity per developer over time, which estimates the effect of the design change with a treated variable and coefficient. The corresponding linear model is 
\[y = \beta_0 + \beta_1 \cdot x + \beta_2 \cdot T\]
where $y$ denotes the ratio of weekend activity for a developer in week $x$ and $T$ represents the treated variable, with $T = 0$ if $x$ is before the change, $T = 1$ otherwise. We fit our model to the data using the python module RDD \footnote{\url{https://GitHub.com/evan-magnusson/rdd} (February 23, 2020)}. We fit the model with several bandwidths, denoting the number of weeks we consider in total before and after the week of the design change. 

We report the results in Table~\ref{tab:weekendRDDstats}. Whether we consider 1, 2, or 3 weeks before and after the design change (corresponding to bandwidth values of 2, 4, and 6, respectively), we find that there was a significant decrease in the number of contributions made on weekends following the change. 

\begin{table*}[htbp]
\begin{tabular}{|l|l|l|l|l|l|}
\hline
\textbf{Bandwidth} &
\textbf{\# Obs.} & 
\textbf{$\beta_0$ (intercept)} &
\textbf{$\beta_1$ (x coeff.)} & 
\textbf{$\beta_2$ (treated coeff.)} &
\textbf{p-value $\beta_2$} \\ \hline
 2 & 73433 & 0.0582 & 0.0358 & -0.0985 & $<0.001$ \\ \hline
 4 & 144726 & 0.1843 & 0.0050 & -0.0365 & $<0.001$ \\ \hline
 6 & 214249 & 0.2016 & 0.0004 & -0.0241 & $<0.001$ \\ \hline
\end{tabular}
\caption{Regression discontinuity design model estimates of the change in the share of activity carried out on the weekend. The bandwidth column considers the number of weeks considered in total before and after the design change.}
\label{tab:weekendRDDstats}
\end{table*}

\begin{figure}[!]
\centerline{\includegraphics[width=.5\textwidth]{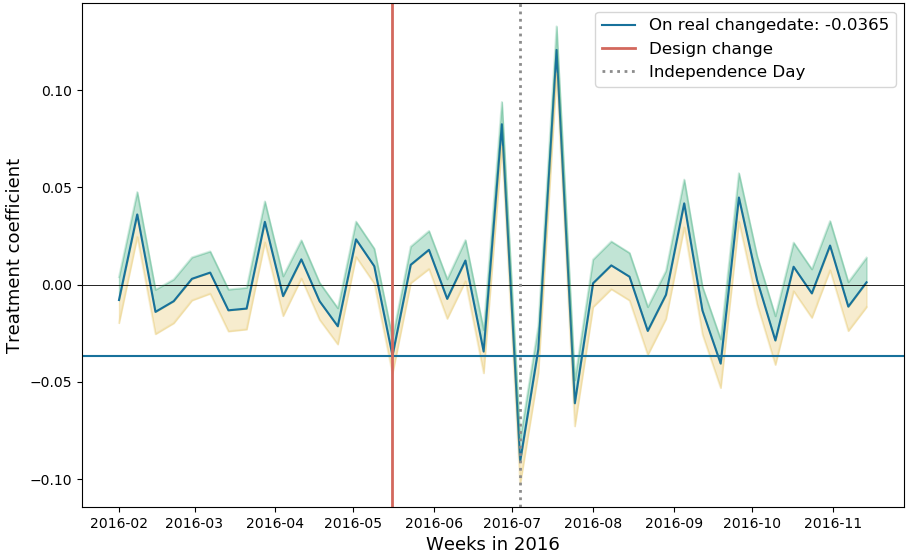}}
\caption{Results of the RDD weekend placebo test with fake change dates and bandwidth 4. The estimated treatment of the actual design change on the share of weekend contributions is highlighted in red. All other points represent the result of repeating the same analysis with a hypothetical design change in other weeks. The green and yellow shaded regions represent the 95\% confidence intervals around the estimated coefficient. Prior to the design change, no other week saw such a large estimated treatment effect as the actual design change week.}
\label{fig:weekendPlaceboRDD}
\end{figure}
In order to test the robustness of our findings, we carried out a series of tests using the same model with the design change artificially set to different dates in 2016. Such tests are known as placebo tests in the econometrics literature~\cite{imbens2008regression}. To keep results comparable, we again only focus on active developers with at least 30 contributions in the respective time interval and use bandwidth 4. Figure~\ref{fig:weekendPlaceboRDD} shows the resulting treated coefficients for all tests with the placebo date in week $x$. Before the change, we observe no higher treated coefficient than the original one of -0.0365 and larger 2.5\% confidence intervals in general. After the change we observe fluctuating coefficients around the Independence Day but also similar values in September and October. We make two points here: first, the fall in weekend work around the fourth of July weekend seems to be compensated by overwork on neighboring weekends. Second, all placebo points before the design change show no difference as large as the real design change, suggesting that this was in fact a significant change.

\subsubsection{Single Contribution Days}

\begin{figure}[ht!]
\centerline{\includegraphics[width=.5\textwidth]{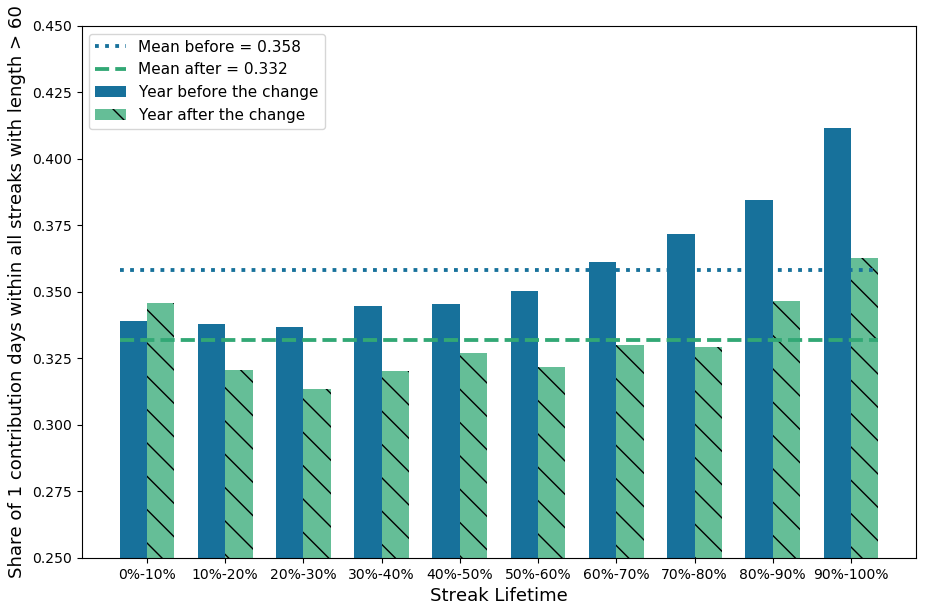}}
\caption{Share of days with one contribution by decile over all streaks one year before and after the change with a minimum streak length of 60. Single Contribution Days are not uniformly distributed and have a higher share at the end of a streak. After the change this tendency weakens.}
\label{fig:OCD}
\end{figure}

Besides steering users to make contributions on weekends, the counters likely exerted significant pressure on users with long ongoing streaks. If this is the case, we expect that users on long streaks before the change are significantly more likely to have days in
which they do the minimum activity to extend their streak. We call such days Single Contribution Days (SCD). In Figure~\ref{fig:OCD}, we plot the distribution of SCDs by streak-length decile in streaks of length 60 or higher before vs. after the change. We see that SCDs were more common before the change overall (36\% of days vs 32\%). At the end of long streaks before the design change, over 40\% of days were SCD, compared with roughly 36\% after the change. We interpret this as evidence that developers went out of their way to keep their long streaks alive.

\subsection{RQ3: Counters for Goal-Setting}
We have seen evidence that many developers stopped streaking after the counters were removed, and that this reflected changing patterns of contribution. These findings suggest that developers were interested in the value of the counter for signaling purposes. Another possible motivation for engaging with the counters was that they could help developers set and stick to long term goals. Indeed there are many resources offering to guide a learner to a goal through a program of daily activity. One example in the world of computer programming is the ``100DaysOfCode'' challenge~\footnote{\url{https://www.100daysofcode.com/}}. As the name suggests, the challenge's goal is to code at least one hour a day for 100 days in a row. Participants are encouraged to fork a GitHub repository, which serves as a journal template and can be filled with daily individual progress updates. Though these daily journal updates do not count as a valid contribution for the streak counters as they are done in a forked repository, we assume that the population of developers forking this repo is significantly more likely to engage in streaking behavior with goal-based motivations. By observing this population across the change, we can check whether the counters improved the chances that developers would achieve their goal.

\begin{figure}[t]
\centerline{\includegraphics[width=0.5\textwidth]{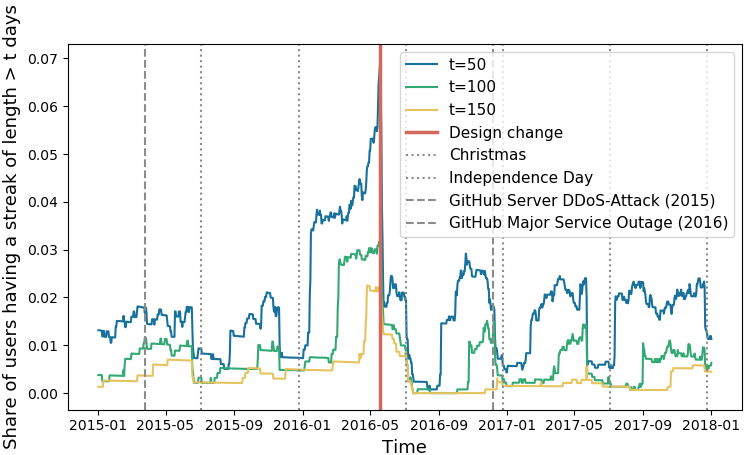}}
\caption{The streaking behavior of developers who forked a code 100 days in a row-type GitHub project. The largest drop in streaking behavior occurred immediately after the design change on GitHub (red line). Share of developers having streaks above the 100 days goal decreased after the change permanently.}
\label{fig:shareGOAL}
\end{figure}

First we used the GitHub API to search for further goal based communities on GitHub with similar 100 days of contributions goals. The data we share online includes a list of the projects we found. We also used the API to collect developers forking the corresponding template repositories and translates their usernames to IDs in our database. From a collection of roughly 16,000 developers forking any one these projects, we found more than 1,600 developers in our filtered data (recall that our filtered dataset only contains developers for which a location could be inferred). Figure~\ref{fig:shareGOAL} shows the daily share of these developers over time having a streak of length $t$ for $t=50,100,150$. In the year before the change, we observe a sharp increase in streaking with several drops (most likely caused by developers who reached their goal). But we also observe streaking beyond 150 days. Within the days after the change, a large amount of developers stopped streaking immediately for all thresholds $t$. Developers seem to be discouraged by the design change and gave up their goal. However, we observe surviving and new streaks longer than 50 days after the design change in 2017, but with less developers participating compared to the year before. The share of developers streaking above their goal over 150 days decreased permanently. What is unclear from this figure is whether the developers who are still streaking after the change reach their goal of 100 days.

\begin{figure}[t]
\centerline{\includegraphics[width=0.5\textwidth]{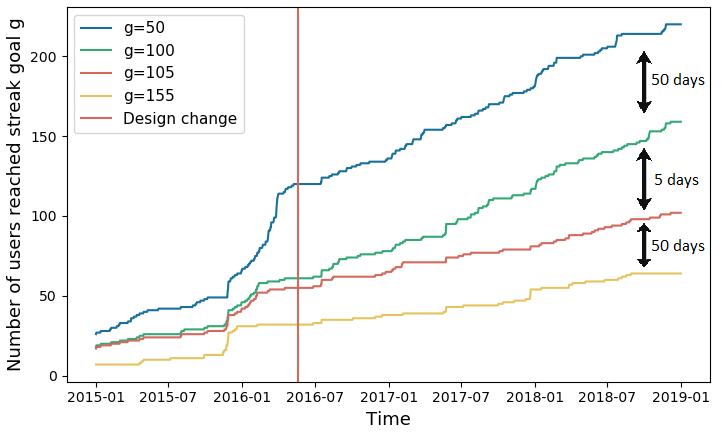}}
\caption{The streaking achievements of developers who forked a code 100 days in a row-type GitHub project. Number of developers reaching streaks of length g days for g $\in \{50,100,105,155\}$ over time. The growth of g=50 achievers decreases after the change, but developers still hit new goals. The large difference between achievers of g=100 and g=105 emphasizes anchoring effect of the 100 day goal.}
\label{fig:goalLifetime}
\end{figure}

Figure~\ref{fig:goalLifetime} plots the total number of developers achieving streaks of length $g \in \{50, 100, 105, 155\}$ over time. The design change seems to have a low impact on these statistics, as many developers still achieve significant streak lengths after the change. But when comparing differences between the number of achievers of different goals, we observe nearly the same increasing gap between achievers of the 50 day/100 day real goal and achievers of 100 day/105 day streaks. Thus, many developers stopped maintaining their streaks right after hitting the 100 day goal, not even reaching a length of 105 days. These results suggest that developers still streak because of the goal based challenge after the change, even without having a streak feature. Moreover, the forked journal with daily updates could have helped to keep track of a streak, as many developers stop streaking quite exactly after reaching a length of 100 days. This suggests that some developers did not need the counters to achieve their goals.

\subsection{RQ4: Imitation in the Social Network}
GitHub, like many other online platforms for collaborative work, includes a social network. Developers can follow each other and receive updates about the activities of their network neighbors. It is likely that developers visit the profiles of their friends in the social network more often than those of other developers, and so were more likely to observe the streak counters of their friends. In this section we ask whether there is any evidence that developers imitated their neighbors in streaking behavior. Observing such a peer effect would demonstrate that gamification can modify developers behavior through social networks. Again we exploit the site design change: we compare the correlation of streaking behavior of network neighbors before and after the removal of counters. 

Why do we expect that gamification influenced behavior through the social network? Observing the performance of familiar others can inspire people to try harder. In fact, in seeking to evaluate and benchmark our own performance, we often seek out information about others~\cite{festinger1954theory}. In the context of gamification, in which points or badges may seem arbitrary, a relative comparison seems essential to define the value of rewards. Previous studies do find significant correlations in engagement with gamification between developers who are connected and can view each others' outcomes~\cite{hamari2015working}. It is often unclear, however, whether these correlations are due to sorting or influence. 

Sorting, sometimes called homophily, refers to the phenomenon that similar individuals are more likely to become friends. In the case of GitHub, developers may be more likely to connect with developers with similar work schedules or degree of motivation. This latent similarity would explain similar degrees of streaking behavior among connected developers. Influence, on the other hand, refers to the tendency of friends to become more similar over time, whether because of imitation, a desire to conform, or other social forces. Developers on GitHub may be influenced by the activity patterns of their neighbors. In the case of streaking behavior we suggest that such influence was likely enhanced by the counters present on developers profiles before the design change.

In general these two factors are confounded when conducting observational studies~\cite{shalizi2011homophily}. For example, we cannot easily tell if two connected developers both have high streak scores because they are influencing one another to work harder, or if they connected in the first place because they are similarly dedicated to working. Yet the removal of the streak counters presents an opportunity to partially disentangle these effects, if it is interpreted as a natural experiment~\cite{aral2017exercise,ternovski2020social}. Indeed, the removal of the counters likely suddenly blocked a channel of influence between developers. If streaking is a behavior transmitted by social influence, we would expect the correlation of streaking behavior to fall significantly after the change. 

\begin{figure}[ht!]
\centerline{\includegraphics[width=.5\textwidth]{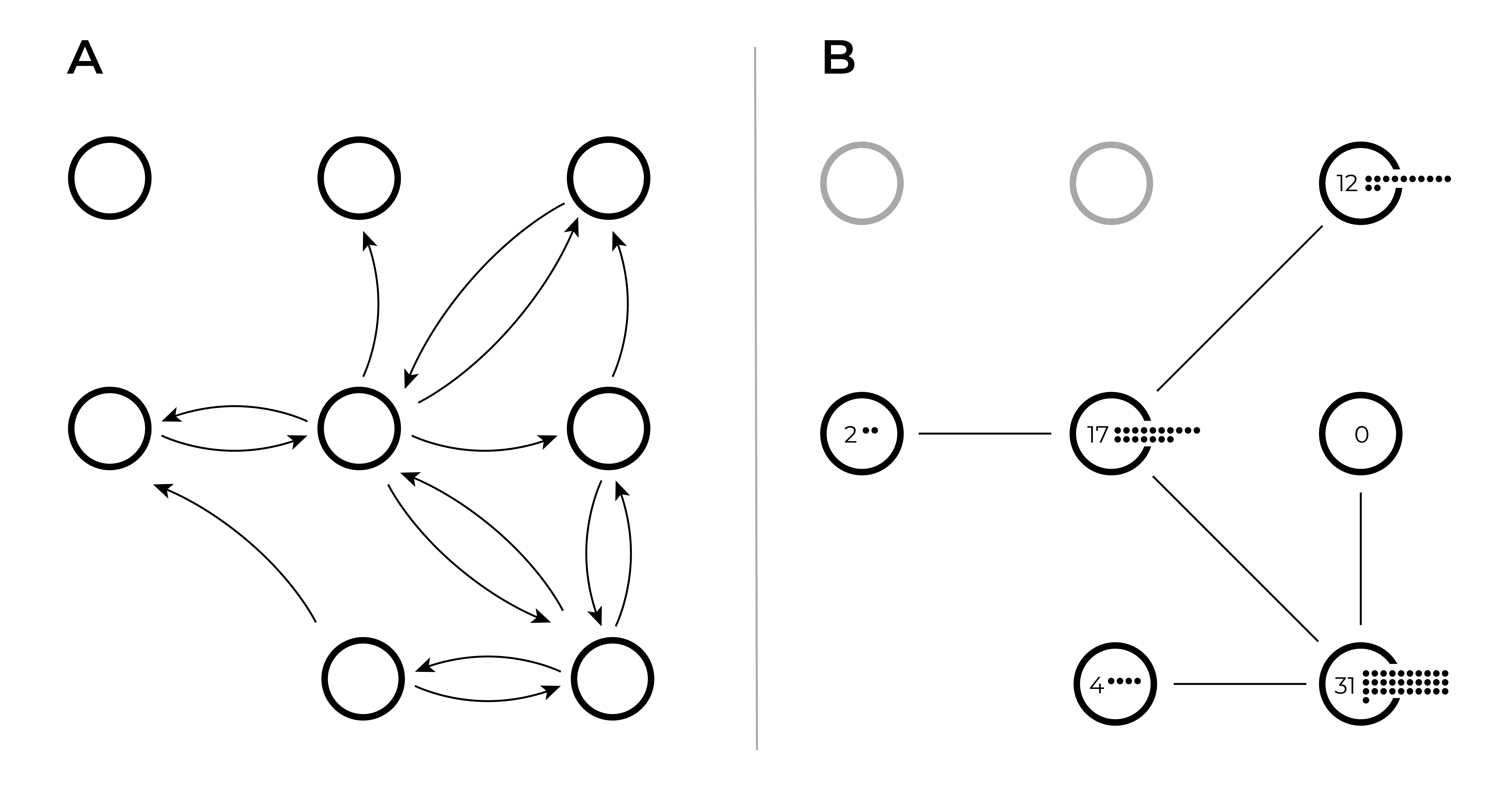}}
\caption{This figure illustrates how we construct a network of social connections of developers on GitHub to study network correlations in streaking behavior. A) Developers on GitHub can follow other developers to stay informed about their activity. We represent developers as nodes and their following relationships as links in a network. Some following relationships are mutual. B) We keep only the mutual connections as they are most likely to represent connections between peers and acquaintances. For any given day, we annotate each developer with the length of their current ongoing streak.}
\label{fig:network}
\end{figure}

To carry out this analysis, we generated a social network from the time-stamped following relations stored by GHTorrent. In this network, nodes represent developers while links between them are mutual follower connections. Since we want to focus on ties between mutual acquaintances, we do not observe single/non-mutual follows and discarded all nodes with a degree of $0$. We visualize filtering process used to generate the network that we analyze in Figure~\ref{fig:network}. The resulting undirected network snapshot, realized for May 18 in 2016 (the day before the design change) contains 146k nodes and 253k links with an average degree $<k> = 3.46$ and a maximum degree of $k_{max} = 2343$. The network is divided into 11k components, while the largest component contains more than $81\%$ of all nodes.

We labeled nodes as streakers or non-streakers based on the maximum streak length the developer had attained, for different thresholds $t \in \{8, 15, 32\}$. If the maximum streak length of a developer is at least $t$ days long, the developer is considered to be a streaker, otherwise not. To calculate the correlation of streaking status within the network we calculated Newman's attribute assortativity coefficient~\cite{newman2003mixing}). Generally speaking, the assortativity of a network $r$ is a real number in $[-1,1]$, increasing when neighbors tend to have similar attribute values. To analyze the statistical significant of streaking assortativity in the network, we repeat the calculation of assortativity on 1,000 copies of the network in which the streaking label is randomly shuffled. This randomization preserves the network structure and overall prevalence of streaking. We calculate a z-score comparing the observed assortativity with the distribution of assortativity in the randomizations. As an alternative measure of the tendency of streaking developers to be connected, we also calculate a conditional probability: $P(SN | S)$ = $P($node $n$ has streaking neighbor $|$ node $n$ is streaker$)$.

\begin{table*}
\begin{tabular}{c|c|c|c|c|c|}
\cline{2-5}
\textbf{Before Change} (2016/05/18) & \multicolumn{2}{c|}{\textit{Observed network}} & \multicolumn{2}{c|}{\textit{Streak-randomized networks (avg. of 1k)}} \\ \hline
\multicolumn{1}{|c|}{\textbf{Streaking Threshold (Days)}} & \multicolumn{1}{c|}{\textbf{Streaking assortativity}} &
\multicolumn{1}{c|}{\textbf{$P(SN | S)$}} & \multicolumn{1}{c|}{\textbf{Streaking assortativity}} &
\multicolumn{1}{c|}{\textbf{$P(SN | S)$}} &
\textbf{Z-score (assortativity)} \\ \hline
\multicolumn{1}{|c|}{8} & 0.089 & 0.386 & -0.0001 & 0.179 & 41.6 \\ \hline
\multicolumn{1}{|c|}{15} & 0.079 & 0.340 & -0.0001 & 0.082 & 36.9 \\ \hline
\multicolumn{1}{|c|}{32} & 0.045 & 0.253 & -0.0001 & 0.027 & 21.2 \\ \hline
\end{tabular}
\caption{Network assortativity and the conditional probability that a streaker node has a streaking neighbor for the empirical social network on 2016/05/18. We compare these values against their average values under 1,000 randomizations of the streaking label. Both empirical differ significantly from the random experiments at all three thresholds we consider, indicating a connection between streaking and the network structure.}
\label{tab:network2016}
\end{table*}

\begin{table*}
\begin{tabular}{c|c|c|c|c|c|}
\cline{2-5}
\textbf{After Change} (2017/05/20) & \multicolumn{2}{c|}{\textit{Observed network}} & \multicolumn{2}{c|}{\textit{Streak-randomized networks (avg. of 1k)}} \\ \hline
\multicolumn{1}{|c|}{\textbf{Streaking Threshold (Days)}} & \multicolumn{1}{c|}{\textbf{Streaking assortativity}} &
\multicolumn{1}{c|}{\textbf{$P(SN | S)$}} & \multicolumn{1}{c|}{\textbf{Streaking assortativity}} &
\multicolumn{1}{c|}{\textbf{$P(SN | S)$}} &
\textbf{Z-score (assortativity)} \\ \hline
\multicolumn{1}{|c|}{8} & 0.091 & 0.265 & -0.0001 & 0.140 & 37.7\\ \hline
\multicolumn{1}{|c|}{15} & 0.055 & 0.203 & 0.0001 & 0.047 & 22.6 \\ \hline
\multicolumn{1}{|c|}{32} & 0.021 & 0.112 & -0.0001 & 0.011 & 8.3 \\ \hline
\end{tabular}
\caption{Calculations repeated for the empirical network on 2017/05/20, one year after the removal of streak counters from GitHub. We observe smaller but still significant clustering by streaking users, compared with the randomizations.}
\label{tab:network2017}
\end{table*}

Table~\ref{tab:network2016} shows that the streaker attribute is not randomly distributed, as we observe an assortativity around 0 for the randomized networks and between $0.04$ and $0.09$ for the real network. We calculated a z-score to test the statistical significance of the difference in assortativity between the empirical graph and the random simulated graphs, with $z \gg 1.96$ for all tests. Values decrease with an increasing streaker threshold $t$, since there are fewer streakers and remaining streaking nodes have a higher fraction of non-streaking neighbors. The conditional probability also suggests that there is significant clustering of streaking developers in the network. The probability that a streaker is connected to another streaker is 38.6\% compared to the average of 17.7\% in the randomizations for $t=8$. With an increasing $t$ the difference between the networks increases too. At $t=32$ we observe $P(SN | S)$ = 25.3\% compared an average of 2.8\%  in the random networks. 

This suggests that streaking developers are significantly interconnected, but does not tell us whether sorting or influence are at play. If we repeat the analysis of homophily among streakers after the change, we can test these factors. If only sorting is at play, i.e. if individuals are more simply likely to connect with people who have the same tendency to streak, there should be no change in the observed assortativity levels. If only influence is present, then there should be little or no assortativity remaining after the counters are removed. Results in between suggest that both effects were present before, and that the design change blocked an important channel for influence. Indeed this is what we hypothesize: that some developers were driven to extend their streaks because they observed higher totals among their neighbors in the network, and that the design change ended this phenomenon. 

We thus created the same network one year after the change and only observed streak records in these 12 months for the streaker attribute. Besides 20k new existing nodes, there is an overall increase of 13.4\% in the number of edges and a general increase in connectivity among nodes. Table \ref{tab:network2017} shows that streaking in the network remains assortative and that post-change streakers are still connected. But the overall values decreased significantly compared with the random networks, relative to what we observed in the network from 2016. Notably smaller z-scores suggest a weaker level of streaking assortativity after the change. For the threshold $t=8$ only every fourth developer is connected to another streaker, while one year before we observed 38.6\%. This suggests that the signals provided by the streak counters were indeed a conduit for peer effects in the social network of GitHub developers. In other words, it appears that the counters spurred developers to keep up with or exceed the streaks of their neighbors in the social network. The remaining assortativity can likely be attributed to sorting.

\section{Discussion}
Our use of a natural experiment on a widely-used online platform offers a new perspective on some the main issues of gamification research today~\cite{koivisto2019rise}. Methodologically, our results are based on data on the scale typical of observational studies, while retaining some flavor of an experimental study (for instance that we can exclude several confounding explanations for our findings such as a secular change in behavior). Theoretically, our focus on the context of software development on GitHub gives us a clearer lens through which to interpret the interaction between developer and gamification. Our findings should give pause to decision makers considering whether to implement gamification, especially in software~\cite{garcia2017framework}.

What lessons can platform designers, in particular those designing for software developers, draw from our study? The first is that user responses to gamification can be highly varied. We speculate that some users respond to gamification because they would like to signal status or commitment. Others may use gamified elements to set and stick to goals. Yet others may learn behavior or even evaluate themselves by comparing their gamified achievements with those of their friends and collaborators. In sum, any game designer must consider that users may engage with new games in unexpected ways. In particular, some users may focus their efforts on collecting points and badges to the detriment of the actual content of their activity. This is worth keeping in mind even for designers who seek to tweak systems and platforms to virtuous ends~\cite{grigoreanu2008can,johnson2016gamification}. The observed effects of the removal of the counters implies that platform designers have some responsibility to consider how the introduction of gamification elements steers behavior.

Indeed, some users may chase the rewards of gamification to an unhealthy degree. Long streaks of uninterrupted contributions may lead to burnout. Indeed, some emotional responses to GitHub's announcement that the streak counters were no longer part of developer profiles suggest that some developers had developed an unhealthy relationship with these elements~\cite{blogRemoving}. It also seems to us unlikely that developers logging in to make a single contribution to maintain an ongoing streak made useful or high quality contributions. This sort of behavior reflects an optimization of individual behavior for the sake of the game, and not for the quality of the work. These shortcomings of the counters might have been evident to GitHub's designers, who, after all, removed these features from their platform. Nevertheless, the patterns of behavior we observe could generalize to other platforms and games. As gamification proliferates in online platforms and labor markets, we argue it is important to consider these findings.

\subsection{Limitations}

We now highlight several limitations of our data and technical approach. The streak computation itself is very sensitive to small changes in the source data, as a single missing day in our data would end all streaks immediately. Fortunately we do not observe such patterns at the macro scale. Another assumption that we make about our data is that developers do not frequently edit the contribution time of commits. In this way it was technically possible for developers to create artificial streaks of arbitrary length. Besides filtering out developers that were clearly engaged in such behavior (for example those with commits at times decades before the creation of the GitHub platform), we assume that this behavior was rare. Lastly, a more ideal natural experiment would have observed both the introduction and the removal of the counters.

Considering our data sample, another limitation is that we only consider developers for which GHTorrent could reliably infer location. This is a necessary step to calculate streaks but introduces bias to our sample, as developers who provide information about their location are likely different in motivation, attitude, and behavior from developers who do not. Geolocation inferences are also more accurate for residents of major cities and Western countries~\cite{johnson2017effect}. We also acknowledge that developers may have moved time zones during our period of analysis. 

Finally, we note that even though GitHub removed the streak counters from user profiles in 2016, the colored contribution graph remains a part of profiles to this day. This gamified element gives visitors to a profile an impression of a user's activity over time at a glance. In this way incentives remain to signal consistency of contributions over time, and the calendar offers ways, if more limited than the counters, to track progress towards goals and for collaborators to influence one another.

\subsection{Future Work}

To better understand potential effects of gamification on user behavior, it is important to understand how the gamified element in question taps into different psychological motivations users have. For example, the ability to signal commitment via a high streak counter can be useful for individuals on the labor market, particularly in software development. Yet chasing that signal can lead to bad outcomes via single contribution days or overwork. More research is needed to study how different kinds of gamification (such as counters, leaderboards, or badges) steer behavior by appealing to different motivations~\cite{sailer2017gamification}, for example the desire to signal abilities to others~\cite{easley2016incentives} or to track progress towards a specific goal~\cite{fortes2018theory}. We know, for example, that extrinsic motivators such as points or rewards can ``crowd out'' intrinsic motivations for pro-social behavior in some contexts~\cite{ariely2009doing}. Any analysis of the motivation of users should recognize that the socio-demographic backgrounds and values of users are significantly related to their responsiveness to gamification~\cite{eickhoff2012quality,koivisto2014demographic}.

We have not discussed how gamification elements on online platforms may lead to biased evaluations of its users~\cite{thebault2017toward,thebault2017simulation,hannak2017bias,may2019gender}. Individual characteristics of users, such as their gender, ethnicity, and cultural origins undoubtedly relate to their propensity to engage with gamification. If gamification rewards are then used to, say, rank top users, this can lead to run-away inequalities in outcomes on a platform. There is a significant potential for such bias when algorithms interact with gamified elements in a complex way~\cite{bokanyi2020ride}. Regarding the field of software development in particular, future work should engage with the literature on engineer productivity to design effective gamification~\cite{meyer2017work}. For instance, the lessons of recent work on goal-setting methods to foster good habits among software developers could be applied to this question~\cite{meyer2019enabling}. 

Our paper has also focused, to a large extent, on the responses of individuals to gamification. Yet gamification is generally employed with the goal to improve communities in some way, and sometimes this is the primary purpose of such features~\cite{irannejad2017sociotechnical}. Future work on the impacts of gamification should zoom out from the individual to study collective outcomes, for example if projects or communities that engage significantly with gamified elements perform better~\cite{repobadges}.

\subsection{Conclusion}
In this paper we presented evidence from a natural experiment that gamification steers behavior and increases participation among software developers, though potentially in undesirable ways. We urge the designers of online platforms to consider the potential consequences of adding gamification elements to their sites. Our findings suggest that some users will change their behavior to collect digital tokens, but that this behavior may optimize for the game, and not necessarily for healthy and productive activity.

\section*{Acknowledgements}
We thank Ancsa Hannak, Srebrenka Letina, Theresa Gessler and Zsofia Czeman for helpful discussions and feedback. This paper is based on the bachelor thesis of the first co-author at RWTH Aachen University.

\bibliographystyle{IEEEtran}

\bibliography{sample-base}

\end{document}